\documentclass[twocolumn,11pt]{article}
\usepackage{natbib}
\usepackage{amsmath, amssymb}
\usepackage{graphicx}
\usepackage{cases}
\usepackage{aas_macros}
\usepackage{times}
\usepackage{url}

\begin{document}

\title{Classification of exoplanets according to density}
\author{
Andrzej Odrzywolek$^{1,2\ast}$
and 
Johann Rafelski$^{1}$\\[0.3cm]
\normalsize{$^{1}$Department of Physics, The University of Arizona 
Tucson, AZ 85721, USA}\\
\normalsize{$^{2}$M. Smoluchowski Institute of Physics, Jagiellonian University Cracov, Poland}\\
\normalsize{$^\ast$To whom correspondence should be addressed; E-mail: andrzej.odrzywolek@uj.edu.pl\;.}
}

\date{}

\maketitle 

\begin{abstract}
Considering probability distribution as  a function of the average density $\bar{\rho}$ computed  for 424 extrasolar planets we identify three log-normal Gaussian population components. The two most populous components at $\bar{\rho}\simeq0.7$~g/cc and $\bar{\rho}\simeq7$~g/cc are  the   ice/gas giants and iron/rock super-Earths, respectively. A third component at $\bar{\rho}\simeq30$~g/cc is  consistent with  brown dwarfs, {\it i.e.}, electron degeneracy supported objects. We note presence of several extreme density planetary objects. 
\end{abstract}

\paragraph*{The raw radius-mass data.}
Our objective is to recognize statistical regularities and possible anomalies in the physical state of the matter according to density \citep{1975Sci...187..605W} addressing the databases  of exoplanets \citep{2016Natur.530..272H,2015Natur.527..288W,2014Natur.513..336L}. The average density $\bar{\rho}$ of the planets 
\begin{equation}
\label{rho}
\bar{\rho} = \displaystyle\frac{M}{\frac{4}{3}\pi R^3}\;,
\end{equation}
is closely related to the theoretical mass-radius ($M-R$) relation \citep{2013Sci...340..572H}.

Source of the data is the NASA exoplanet archive, \url{exoplanetarchive.ipac.caltech.edu} \citep{NASAdatabase} and The Extrasolar Planets Encyclopedia, \url{exoplanet.eu} \citep{2011A&A...532A..79S}. Both were retrieved on 22 October 2016. Number of objects reported with both $M$ and $R$ is 510 out of 3388 and 610 out of 3533, respectively. To ensure quality of data we concatenated databases, merged duplicates and split into \lq\lq gold\rq\rq, \lq\lq silver\rq\rq\ and \lq\lq bronze\rq\rq\ subsets.  The \lq\lq gold\rq\rq\ sample of 424 includes only exoplanets data with consistent (but not necessarily identical) and unambiguous values $M,\;R$  in both sources, and reviewed in original sources \citep{Kepler-128-TTV-mass,TTV_densities,Kepler-37b-mass-source} all dubious cases. \lq\lq Silver\rq\rq, including 146 objects, includes unconfirmed results appearing only once, and the remaining \lq\lq bronze\rq\rq\ data includes $\sim100$ upper mass limits only. 

\begin{figure*}[t]
\centering
\includegraphics[width=0.94\textwidth]{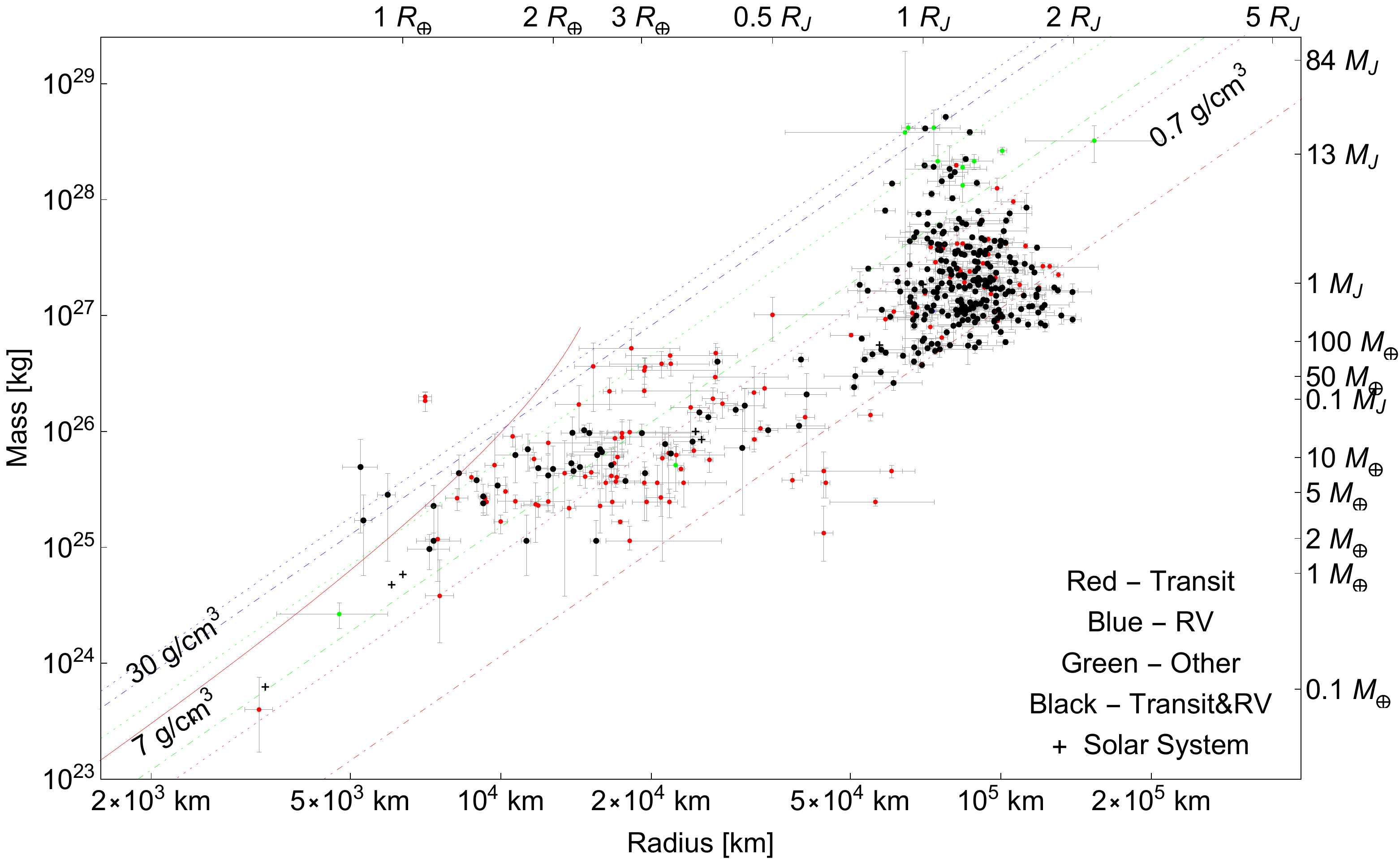}
\caption{\label{MassRadius} Scatter plot in mass  -- radius plane (log scale) of raw data for 432 (exo)planets. Data points are color coded according to detection method: red: transit; blue: radial velocity (RV); green: imaging, microlensing; black: both RV and transit. Diagonal lines along constant average density   delimit $1\sigma$-domains identified in our analysis as belonging to the three main families of exoplanets, see text.  
}
\end{figure*}

\begin{figure*}[h]
\centering
\includegraphics[width=0.94\textwidth]{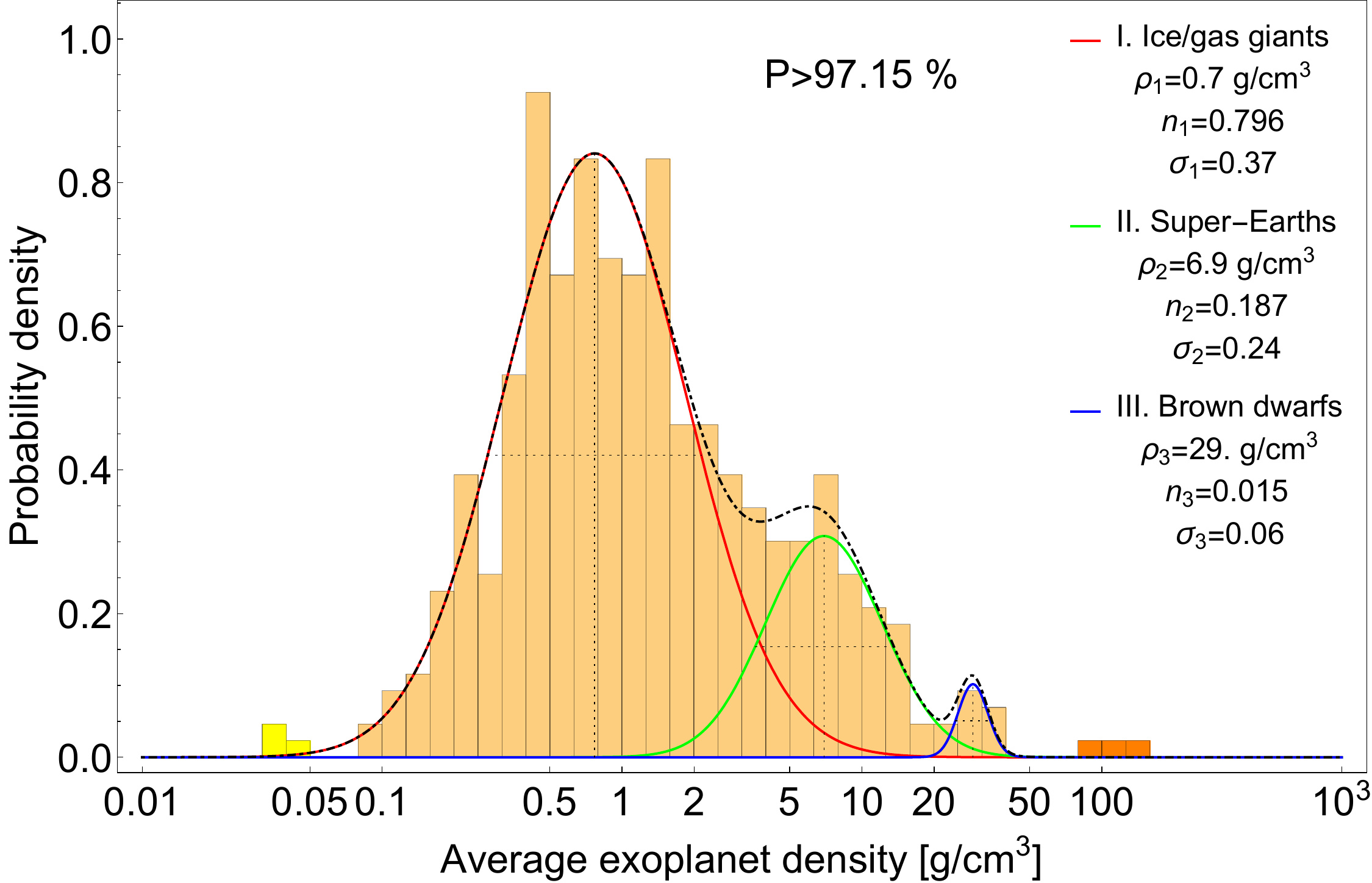}
\caption{\label{fit3peaks} Distribution of the average density for 424 exoplanets and   8 Solar System planets. Histogram of 32 data bins is shown for visualization purpose only; our fit uses the density data directly. Interpretation/names of the component curves is based on  the study of average density only.
}
\end{figure*}

In this analysis only the \lq\lq gold\rq\rq\ sample plus eight Solar System planets were used. The here considered raw  $M$--$R$ data is presented visually in Fig.~\ref{MassRadius}. Curved (red) line shows the theoretical radius-mass relation for a pure Fe planet \citep{2013PASP..125..227Z}. Solar System planets are marked by $+$. The resulting histogram for base-10 logarithm of the density is shown in Fig.~\ref{fit3peaks}, using 32-bins chosen for visual convenience.


Three low density outliers below 0.05~g/cc (yellow bars in Fig.~\ref{fit3peaks}: Kepler-51 b,c,d \citealt{Kepler-51-inflated}) and three high density above 50~g/cc (orange bars in Fig.~\ref{fit3peaks}: Kepler-128 b,c \citealt{Kepler-128-TTV-mass}, Kepler-131 c \citealt{Kepler-37b-mass-source}) are also visible  in Fig.~\ref{MassRadius} below and respectively, above the diagonal lines. These are separated from the bulk of data and are excluded from the statistical analysis. 

\paragraph*{Data analysis.}
Wolfram \textit{Mathematica 11}  \pmb{EstimatedDistribution} command \citep{wolfram} was used to process our data set of 418 exoplanets (424 less 6 outliers) + 8 Solar System planets. Log-likelihood maximum for the data is the continuous black dot-dashed curve shown in Fig.~\ref{fit3peaks}. This result suggests that the probability density for exoplanet data $\bar{\rho}$ is the superposition of three log-normal Gaussian distributions.

The two biggest Gaussian components in Fig.~\ref{fit3peaks} (red and green lines) can be recognized in the density distribution figure before the numerical fit. A third smaller component (blue line) emerges as an additional component during numerical treatment. Values for positions $\bar{\rho}_k$, normalizations $n_k$, and standard deviations $\sigma_k$ of the three Gaussians are shown within contents off Fig.~\ref{fit3peaks}. The Pearson $\chi^2$ probability test shows a value above $P>97.15\% $ for the 3-Gaussian probability density function shown in Fig.~\ref{fit3peaks}. 

The envelope curve seen in Fig.~\ref{fit3peaks} for exoplanet data is thus a superposition of three  dimensionless probability distributions
\begin{equation} \label{dimensionless} 
\frac{dP}{d \lg{\bar{\rho}}} = \rho \frac{dP}{d \bar{\rho}}\;. 
\end{equation} 
A probability distribution normal in $\lg{\bar{\rho}}$ could be due to extraneous factors such as data sampling during observation, but it could also be related to scale-free planet formation mechanisms.

\paragraph*{Proposed classification of exoplanets.} 
We classify the populations from left to right in Fig.~\ref{fit3peaks}; the names of the components  are based in our intuitive expectation and prior knowledge. Since distributions overlap objects within a given density range may not be of the same nature.

\begin{enumerate}
\item[I.] {\bf  Ice/gas} 
The first and dominant ($P_I\simeq 80\%$) population depicted as a red line and centered at $\rho_I = 0.7$\;g/cc in Fig.~\ref{fit3peaks} corresponds to the Saturn/Uranus/Jupiter planet type. Considering the full width at half maximum (FWHM), the distribution extends from $\rho\simeq 0.3$\;g/cc to $\rho\simeq 2.1$\;g/cc (Fig.~\ref{fit3peaks}, dotted horizontal segment). Members of this population are found predominantly between red diagonal lines in Fig.~\ref{MassRadius}. 
\item[II.] {\bf Iron/rock} 
The second component ($P_e\simeq 19\%$, $\rho_e=6.9$~g/cc, FWHM from $3.6$ to $13.4$~g/cc, with objects found  between green diagonal lines in Fig.~\ref{MassRadius})  is shown as a green line in Fig.~\ref{fit3peaks}. These objects are very near to Earth\rq s average density of 5.5 g/cc. This is so-called super-Earth population, {\it i.e.}, planets with composition similar to Earth, but often more massive, see e.g. \cite{2015ApJ...811..102P}, Fig.~6.
\item[III.] {\bf Degenerate}
The third and smallest component ($P_d\simeq1.5\%$, $\rho_d \simeq 30$~g/cc, at FWHM extending from $25$ to $34$~g/cc, cf. Fig.~\ref{MassRadius}, blue bands)  is shown as a blue line in Fig.~\ref{fit3peaks}. This density domain overlaps with electron degenerate matter, {\it i.e.} brown dwarfs \citep{RevModPhys.73.719, RevModPhys.65.301}.
\end{enumerate}

Since the three population classes are {\em overlapping} in density, the individual object planet class membership is to be understood in a statistical sense. For example, according to the proposed classification, the Earth, given the average density 5.5~g/cc, has 4.4 times less chance to be an ice giant than super-Earth object. It is possible that with more abundant and precise exoplanet data and allowing for additional information (e.g. range of $M$, and $R$; surface composition) the classification can be made more precise. The super-Earths normalization  ($P_e=19$\%) is smaller than expected based on Solar System experience ($P_e\simeq 50$\%).  This could be result of a bias induced by observational methods available today that favor detection of $M$  and $R$ for large objects as we note visually in  Fig.~\ref{MassRadius}.

\paragraph*{Conclusions.}
Understanding of mean density distribution for exoplanets offers a convenient tool to identify the new and mysterious in the Universe. The knowledge of the widths of the population distributions allows to realize presence of anomalies when larger exoplanet data base becomes available. Our analysis results thus lay out the basis for the discovery of new classes of rare objects, e.g. CUDO \citep{rafelski2013compact}, dark matter \citep{2005Natur.433..389D} or strange matter \citep{1989Natur.337..436S} contaminated exoplanets. Indeed three small ultra-dense outliers are a tantalizing indication of mysteries that the future exoplanet results may reveal.

We proposed that the extrasolar planet distribution is a superposition of three log-normal Gaussians population components allowing the introduction of three classes of exoplanetary objects, distinguishing these by average density. The two classes (I. ice/gas giants, 80\% and II. super-Earths, 19\%) dominate the available data. Our classification in terms of density agrees with the Solar System situation where outer and inner planets are in classes I and II, respectively. The observed relative normalization of the components, strongly favoring the ice/gas class, is probably an observational bias. This bias is also the reason why we do not divide the results seen in  Fig.~\ref{MassRadius} into domains according to $M$-$R$ ranges as the eye easily captures.

The degenerate class  III.  includes about 1.5\% of the available  objects among 432. In mathematical analysis  a separate population can be assigned. On the other hand, 2 x 3 outliers removed from fit are too few to be assigned their own population class. These outliers are inconsistent with the derived probability distribution function (PDF)  at 94\% confidence level in the sense of conservative Pearson $\chi^2$ test, this inconsistency is  higher (97.1\%) if low/high density are considered separately. When more data becomes available it will be possible to decide if the two groups of three density outlier exoplanets are a data fluctuation or, more interestingly,  a new population.  

The proposed classification method employing average density statistics  can be used to analyze growing data sets on other astrophysical objects: stars including white dwarfs, neutron stars, and minor bodies of the Solar System, which analysis is currently underway.

\section*{Acknowledgment.}

A.O. work was supported by The Kosciuszko Foundation. A.O. thanks the Department of Physics, University of Arizona for kind hospitality.

\bibliographystyle{natbib}
\bibliography{ExoDensityShort}

\begin{thebibliography}{}

\bibitem[\protect\citeauthoryear{{Akeson} \& {et. al.}}{{Akeson} \& {et.
  al.}}{2013}]{NASAdatabase}
{Akeson} R.~L.,  {et. al.} 2013, \pasp, 125, 989

\bibitem[\protect\citeauthoryear{Burrows, Hubbard, Lunine \& Liebert}{Burrows
  et~al.}{2001}]{RevModPhys.73.719}
Burrows A.,  Hubbard W.~B.,  Lunine J.~I.,    Liebert J.,  2001, Rev. Mod.
  Phys., 73, 719

\bibitem[\protect\citeauthoryear{Burrows \& Liebert}{Burrows \&
  Liebert}{1993}]{RevModPhys.65.301}
Burrows A.,  Liebert J.,  1993, Rev. Mod. Phys., 65, 301

\bibitem[\protect\citeauthoryear{{Diemand}, {Moore} \& {Stadel}}{{Diemand}
  et~al.}{2005}]{2005Natur.433..389D}
{Diemand} J.,  {Moore} B.,    {Stadel} J.,  2005, \nat, 433, 389

\bibitem[\protect\citeauthoryear{{Hadden} \& {Lithwick}}{{Hadden} \&
  {Lithwick}}{2014}]{TTV_densities}
{Hadden} S.,  {Lithwick} Y.,  2014, \apj, 787, 80

\bibitem[\protect\citeauthoryear{{Hecht}}{{Hecht}}{2016}]{2016Natur.530..272H}
{Hecht} J.,  2016, \nat, 530, 272

\bibitem[\protect\citeauthoryear{{Howard}}{{Howard}}{2013}]{2013Sci...340..572H}
{Howard} A.~W.,  2013, Science, 340, 572

\bibitem[\protect\citeauthoryear{{Lissauer}, {Dawson} \& {Tremaine}}{{Lissauer}
  et~al.}{2014}]{2014Natur.513..336L}
{Lissauer} J.~J.,  {Dawson} R.~I.,    {Tremaine} S.,  2014, \nat, 513, 336

\bibitem[\protect\citeauthoryear{{Marcy} \& {et. al.}}{{Marcy} \& {et.
  al.}}{2014}]{Kepler-37b-mass-source}
{Marcy} G.~W.,  {et. al.} 2014, \apjs, 210, 20

\bibitem[\protect\citeauthoryear{{Masuda}}{{Masuda}}{2014}]{Kepler-51-inflated}
{Masuda} K.,  2014, \apj, 783, 53

\bibitem[\protect\citeauthoryear{{Petigura}, {Schlieder}, {Crossfield},
  {Howard}, {Deck}, {Ciardi}, {Sinukoff}, {Allers}, {Best}, {Liu}, {Beichman},
  {Isaacson}, {Hansen} \& {L{\'e}pine}}{{Petigura}
  et~al.}{2015}]{2015ApJ...811..102P}
{Petigura} E.~A.,  {Schlieder} J.~E.,  {Crossfield} I.~J.~M.,  {Howard} A.~W.,
  {Deck} K.~M.,  {Ciardi} D.~R.,  {Sinukoff} E.,  {Allers} K.~N.,  {Best}
  W.~M.~J.,  {Liu} M.~C.,  {Beichman} C.~A.,  {Isaacson} H.,  {Hansen}
  B.~M.~S.,    {L{\'e}pine} S.,  2015, \apj, 811, 102

\bibitem[\protect\citeauthoryear{Rafelski, Labun \& Birrell}{Rafelski
  et~al.}{2013}]{rafelski2013compact}
Rafelski J.,  Labun L.,    Birrell J.,  2013, Physical Review Letters, 110,
  111102

\bibitem[\protect\citeauthoryear{{Schneider}, {Dedieu}, {Le Sidaner}, {Savalle}
  \& {Zolotukhin}}{{Schneider} et~al.}{2011}]{2011A&A...532A..79S}
{Schneider} J.,  {Dedieu} C.,  {Le Sidaner} P.,  {Savalle} R.,    {Zolotukhin}
  I.,  2011, \aap, 532, A79

\bibitem[\protect\citeauthoryear{{Shaw}, {Shin}, {Desai} \& {Dalitz}}{{Shaw}
  et~al.}{1989}]{1989Natur.337..436S}
{Shaw} G.~L.,  {Shin} M.,  {Desai} M.,    {Dalitz} R.~H.,  1989, \nat, 337, 436

\bibitem[\protect\citeauthoryear{{Weisskopf}}{{Weisskopf}}{1975}]{1975Sci...187..605W}
{Weisskopf} V.~F.,  1975, Science, 187, 605

\bibitem[\protect\citeauthoryear{{Witze}}{{Witze}}{2015}]{2015Natur.527..288W}
{Witze} A.,  2015, \nat, 527, 288

\bibitem[\protect\citeauthoryear{{"Wolfram Research Inc."}}{{"Wolfram Research
  Inc."}}{2016}]{wolfram}
{"Wolfram Research Inc."}, 2016, 'Wolfram|Alpha Knowledgebase"

\bibitem[\protect\citeauthoryear{{Xie}}{{Xie}}{2014}]{Kepler-128-TTV-mass}
{Xie} J.-W.,  2014, \apjs, 210, 25

\bibitem[\protect\citeauthoryear{{Zeng} \& {Sasselov}}{{Zeng} \&
  {Sasselov}}{2013}]{2013PASP..125..227Z}
{Zeng} L.,  {Sasselov} D.,  2013, \pasp, 125, 227

\end{thebibliography}

\onecolumn

\renewcommand\thefigure{S.\arabic{figure}}    
\setcounter{figure}{0}

\renewcommand\thetable{S.\arabic{table}}    
\setcounter{table}{0}

\renewcommand\thesection{S.\arabic{section}}    
\setcounter{figure}{0}

\begin{center}
\Huge Supplement
\end{center}

\begin{section}{Methodology of finding best physics fit}

The sample of 424 available exoplanet mean densities is large enough to exclude most probability distributions but still leaves room for a few good candidates. We find that sum of log-normal distributions (i.e. Gaussians in base-10 logarithm of mean density, $\lg{\bar{\rho}}$) is the  best physics characterization of  the available data. However, even within the log-normal distribution, there is still room for consideration of a variable number  of contributing components. We therefore provide here comparative analysis of the various fits for the density data. An overview of the most important hypothetical distributions considered is presented in Table~\ref{P-value-table}.

To find best parameters, e.g. $\bar{\rho}_0, \sigma$ for the assumed statistical distribution $P(\bar{\rho};\bar{\rho}_0, \sigma)$ the log-likelihood:
$$
\mathcal{L}(\bar{\rho}_0, \sigma) = \sum_{i=1}^N \ln{P_i}, \qquad P_i = P(\bar{\rho_i};\bar{\rho}_0, \sigma)
$$
is maximized. $N$ is the total number of mean density measurements $\bar{\rho_i}$.  Since usually most
probabilities $P_i<1$, $\ln{P_i}<0$ and numerical value for good fits $\mathcal{L} =\mathcal{O}(-N)$, where $N=426$. Better probability density results in less negative log-likelihood, cf. Table~\ref{P-value-table}.

To fit density data, a probability distribution must fulfill some obvious requirements, {\it e.g.} a range from zero to infinity for $ \bar{\rho}$, {\it i.e.} the full real line for $\lg \bar{\rho}$. One often attempts to check first superposition of several Gaussians. However, the main maximum $\bar{\rho} \simeq 1$~g/cc with a width  of the same magnitude and a long tail stretching to high density prohibits use of multiple Gaussian distribution in density, producing P-values which are essentially zero, see last line entry in Table~\ref{P-value-table}.

On the other hand, on first sight a histogram in  $\lg \bar{\rho}$ showed a compact distribution shape with  visible two \lq\lq peaks\rq\rq; that is, two contributing populations.  This observation allows multiple choices for each individual distribution. Therefore we must take into account simplicity of hypothesis and physical intuition to proceed. Arguments strongly favoring simple log-normal, {\it i.e.} Gaussians dependent on  $\lg \bar{\rho}$ as population components, are: (i) density distribution in $\lg{\bar{\rho}}$ is dimensionless; (ii) multiple factors playing a role in planet formation are consistent with the central limit theorem outcome; (iii) automated brute-force fit-search (symbolic regression) places 2-component log-normal mixture distribution on the top of the list as the most likely; (iv) P-values of typical tests, Pearson $\chi^2$ in particular, are the largest we find. 

We see in  Fig.~\ref{fit1peakLogNorm} that a single log-normal population has indeed a small P-value ($P>0.59\%$) while introducing a second log-normal population, see  Fig.~\ref{fit2peakLogNorm}, the  P-value raises to acceptable level ($P>93.5\%$). We thus assume that both population components are described by the same functional  log-normal normal distribution and for comparison purpose we present in Table~\ref{P-value-table} also log-logistic distribution result.

The width of distributions we report has considerable physical significance, and it depends on the possible presence of a third distribution. We thus explore this option further. When adding a third population  we had to select among similar numerical outcome considering three cases: A) A high density third population, a choice motivated by brown-dwarf theories, see Fig.~\ref{fit3peakLogNormA}, which improves the P-value; B) A low density sub-population see Fig.~\ref{fit3peakLogNormB}, which results in reduced  P-value ($P>92.9\%$); and  C) we allow some of the distribution irregularities to be characterized by a narrow distribution, in  exploration of what one calls \lq\lq over-fitting\rq\rq, see Fig.~\ref{fit3peakLogNormC}. Case A (cf.~Table~\ref{P-value-table}) has maximum likelihood, but not   the largest P-value (smallest $\chi^2$) which we find for the case C). Given priors in physics we retain case A) as our choice distribution see Fig.~1 of the main article corresponding to Fig.~\ref{fit3peakLogNormA} of this supplement. However, we keep in mind that there could be still further distinct populations in future data.  Here, we remind that the 2 x 3 data high/low density outliers are not considered in the present analysis and these also could signal additional exoplanet populations.

\begin{table}
\centering
\begin{tabular}{|lc|ccc|}
\hline
distribution  & \#-components  &P-value ($\chi^2$)& log-Likelihood & Figure\\
\hline
log-normal,& 1    &  0.59\% & -343.9 & Fig.~\ref{fit1peakLogNorm}  \\
log-normal,& 2    & 93.47\% & -325.9 & Fig.~\ref{fit2peakLogNorm}  \\
log-normal,& 3-A   & 97.15\% & \textbf{-322.9} & Fig.~\ref{fit3peakLogNormA} \\
log-normal,& 3-B   & 92.86\% & -325.4 & Fig.~\ref{fit3peakLogNormB} \\
log-normal,& 3-C   & \textbf{98.79\%} & -324.7 & Fig.~\ref{fit3peakLogNormC} \\
log-Logistic,& 1  &  0.34\% & -350.9 & -   \\
log-Logistic,& 2  & 84.33\% & -329.8 & - \\
log-Logistic,& 3 & 82.28\% & -327.6 & - \\
normal,& 3   &  $4.8 \times 10^{-6}$\% & - & - \\
\hline
\end{tabular}
\caption{\label{P-value-table} Probability estimates for various log-Normal and other fits.}
\end{table}


\begin{figure}
\centering
\includegraphics[width=\textwidth]{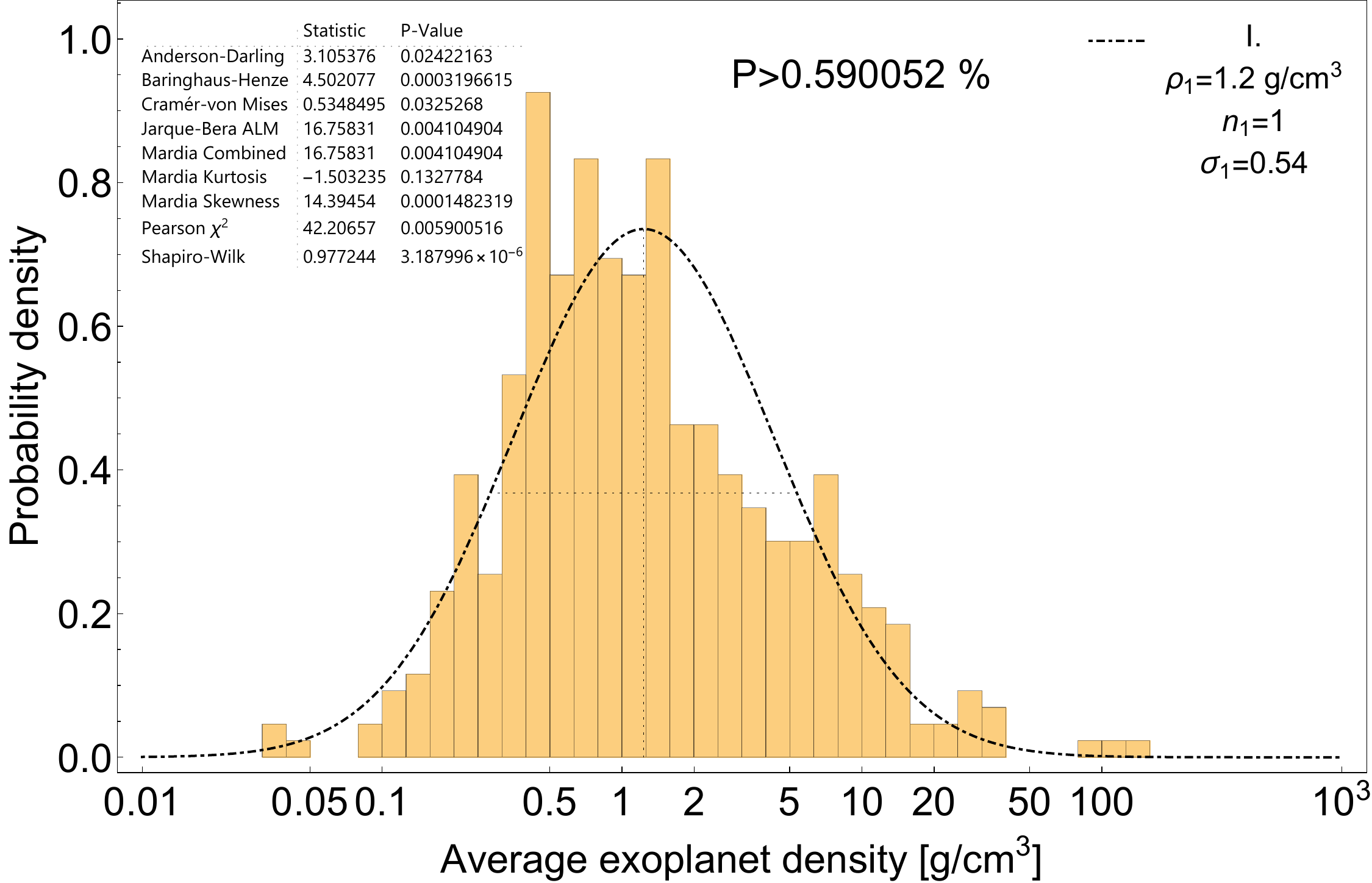}
\caption{\label{fit1peakLogNorm} Single log-normal population fit to the exoplanet density data showing also the resulting P-value and distribution parameters.  
}
\end{figure}

\begin{figure}
\centering
\includegraphics[width=\textwidth]{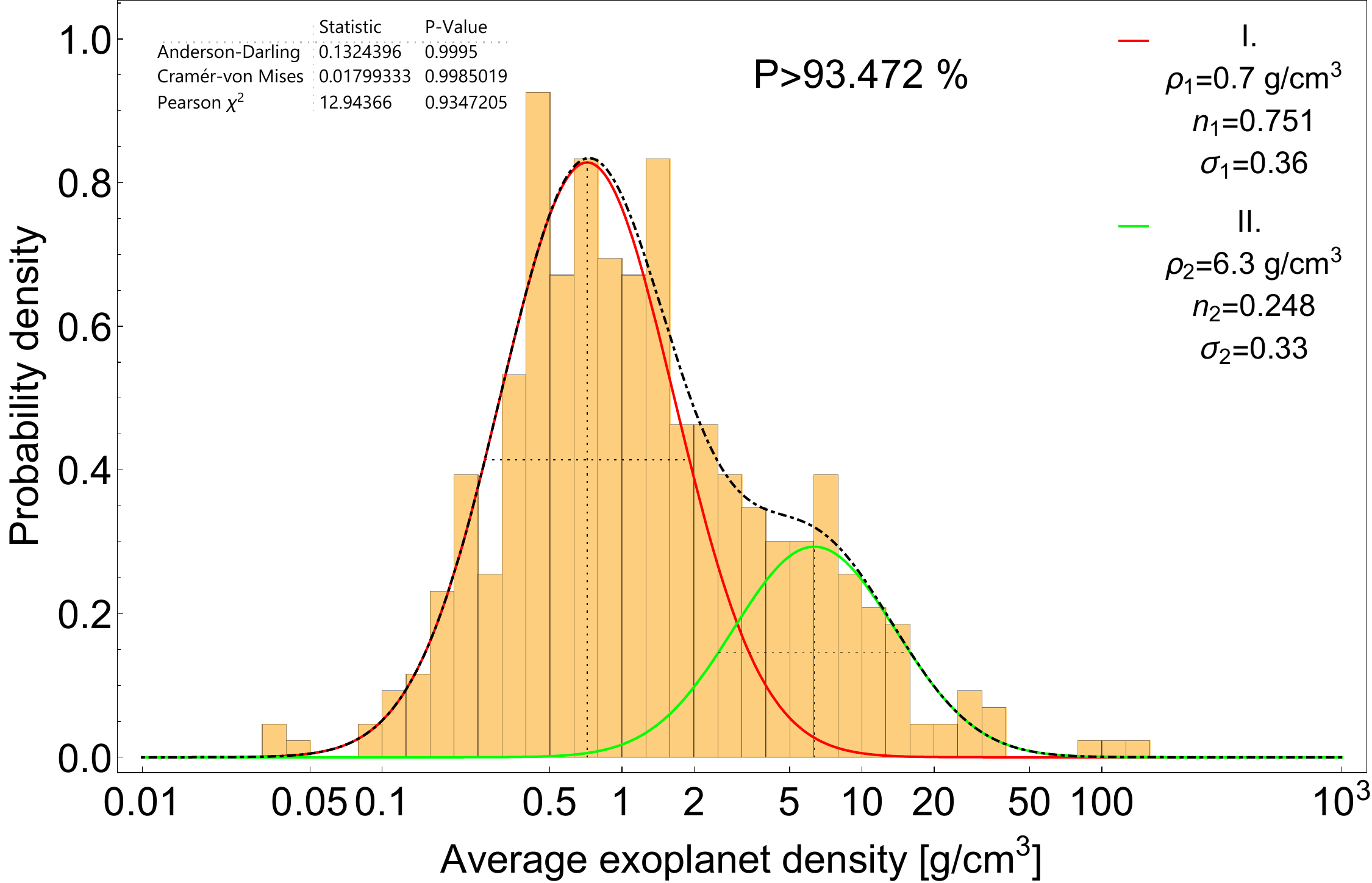}
\caption{\label{fit2peakLogNorm} Double log-normal population fit to the exoplanet density data shewing also the resulting P-value and distribution parameters.  
}
\end{figure}

\begin{figure}
\centering
\includegraphics[width=\textwidth]{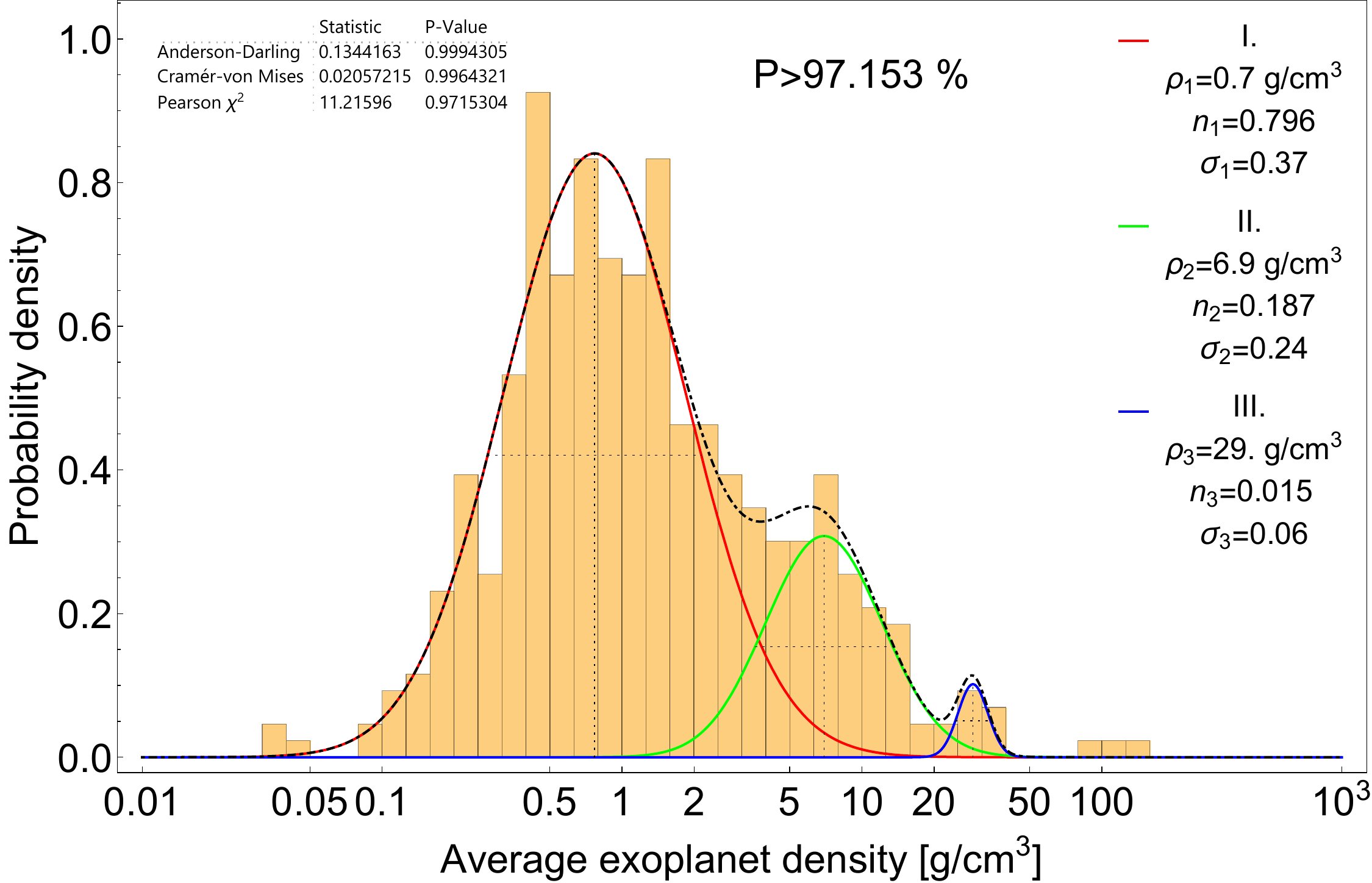}
\caption{\label{fit3peakLogNormA} Case A) triple  log-normal population fit to the exoplanet density data showing also the resulting P-value and distribution parameters.   
}
\end{figure}

\begin{figure}
\centering
\includegraphics[width=\textwidth]{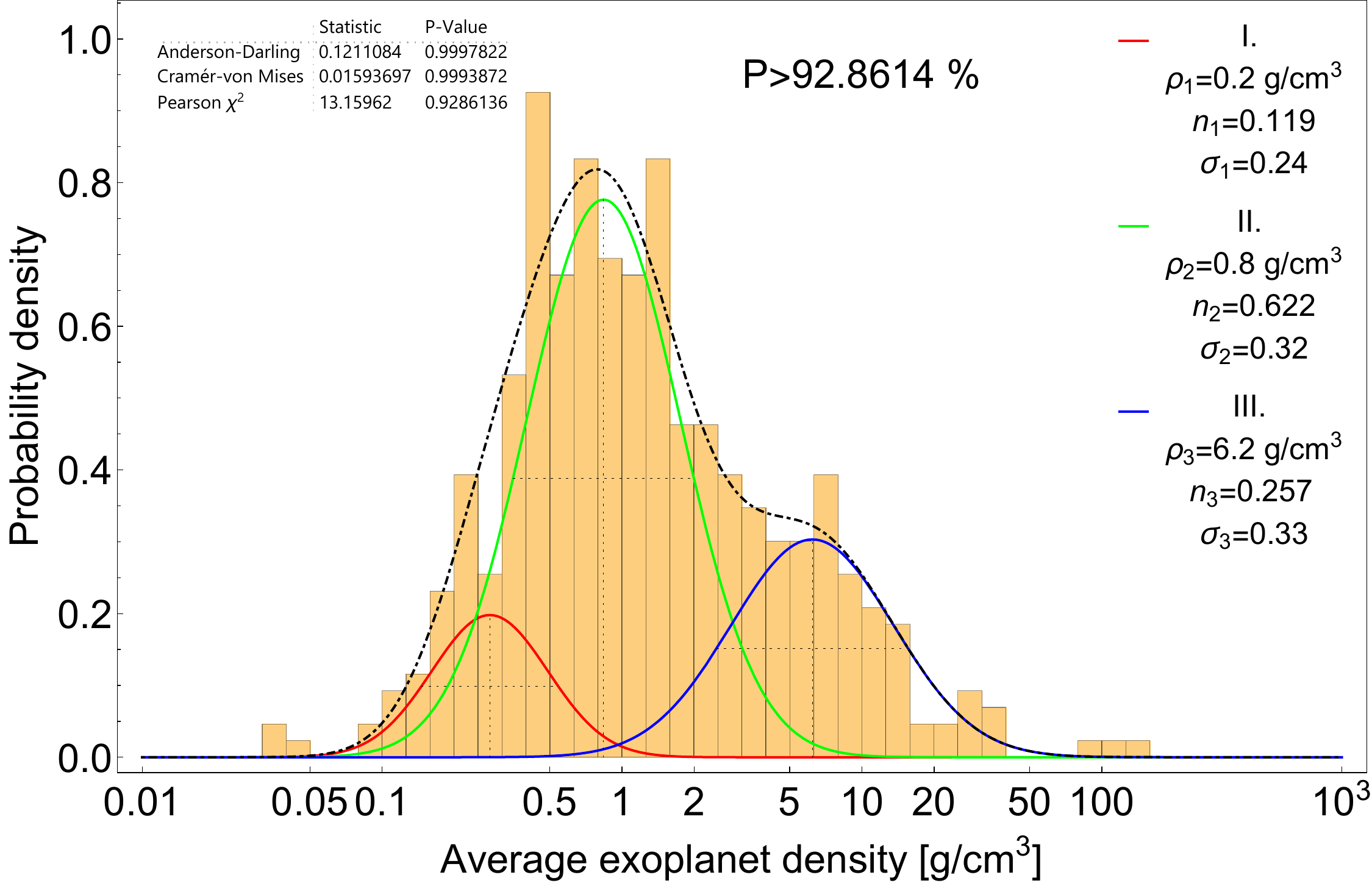}
\caption{\label{fit3peakLogNormB} Case B) triple  log-normal population fit to the exoplanet density data showing also the resulting P-value and distribution parameters.  
}
\end{figure}

\begin{figure}
\centering
\includegraphics[width=\textwidth]{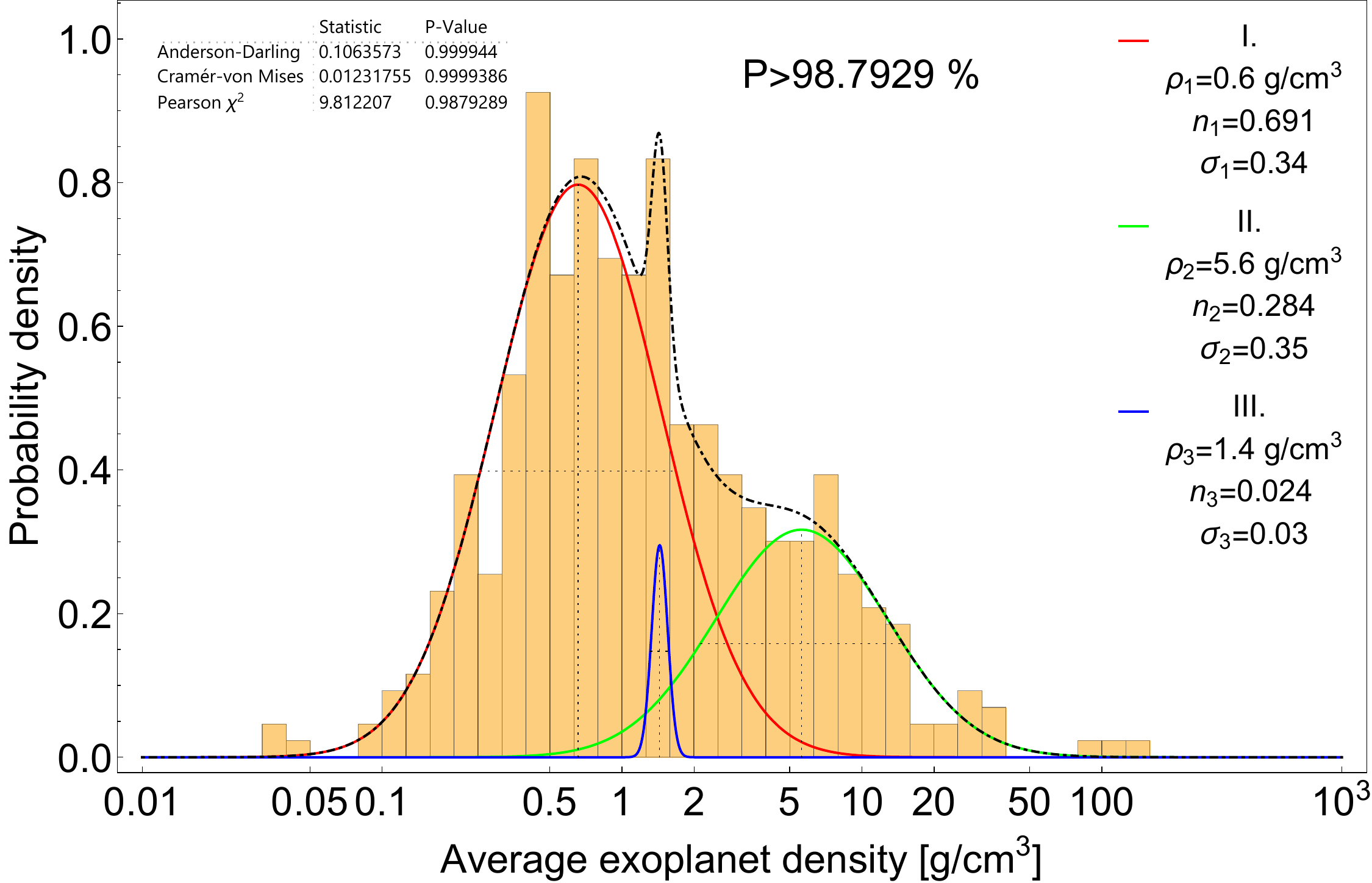}
\caption{\label{fit3peakLogNormC} Case C) triple  log-normal population fit to the exoplanet density data showing also the resulting P-value and distribution parameters with third population focused on $\rho \simeq 1.4$~g/cc.  
}
\end{figure}

\end{section}

\clearpage

\begin{section}{Data files and programs}

Additional files required to reproduce our results are included separately. List and description is provided below.

\begin{itemize}
\item Text file exoplanet\_GOLD.txt (CSV, comma separated values) with 424 gold sample of exoplanets used in analysis.

Header:

{
\tiny 
\verb! Name, SRC, Mneg, M0, M1, M2, Limit, Mass Src, TTV, R, dRneg, dRpos, Name2, RV, Transit, RA, DEC !
}

Column description in file:
\begin{enumerate}
\item Name, selected common name to identify planet
\item SRC, source of the values (NASA or exoplanet.eu)
\item Mneg, max(0,reported mass - error) [Jupiter mass]
\item M0, reported mass [Jupiter mass]
\item M1, reported mass + error [Jupiter mass]
\item M2, reported mass + 2*error [Jupiter mass]
\item Limit, 0/1 mass limit flag
\item Mass Src, method used to measure mass
\item TTV, 0/1 flag indicating use of Time Transit Variation to measure mass
\item R, reported radius [Jupiter radius]
\item dRneg, radius error towards zero [Jupiter radius]
\item dRpos, radius error towards infinity [Jupiter radius]
\item Name2, alternative name
\item RV, Radial Velocity method flag
\item Transit, Transit method flag
\item RA, Right ascension  [degrees]
\item DEC, Declination, [degrees]
\end{enumerate}

\item Mathematica notebook file Exoplanets.nb including code used to obtain results and Figures.

\item Text file MassRadiusFe.dat,  machine readable copy of Table~1 in Ref. LI ZENG AND DIMITAR SASSELOV,
PUBLICATIONS OF THE ASTRONOMICAL SOCIETY OF THE PACIFIC, 125:227–239, 2013
\end{itemize}

\end{section}

\end{document}